\documentclass[aps,prb,amsfonts,amssymb,twocolumn,amsmath,preprintnumbers,floatfix,superscriptaddress]{revtex4-2}%
\usepackage{appendix}
\usepackage{siunitx}
\usepackage{graphicx}
\usepackage{mathrsfs}
\usepackage{bm}
\usepackage{amsmath}
\usepackage{dcolumn}
\usepackage{epstopdf}
\usepackage{dsfont}
\usepackage{amssymb}
\usepackage{tabularx}
\usepackage{longtable}
\usepackage{array}
\usepackage{float}
\usepackage{color}
\usepackage{epstopdf}
\usepackage{mathrsfs}
\usepackage{multirow}
\usepackage[T1]{fontenc}
\usepackage[colorlinks,linkcolor=blue,anchorcolor=blue,citecolor=blue,urlcolor=blue]{hyperref}

\renewcommand{\Re}{\operatorname{Re}}
\renewcommand{\Im}{\operatorname{Im}}
\begin{document}

\title{Extrinsic Contribution to Nonlinear Current Induced Spin Polarization}

\author{Ruda Guo}
\address{Center for Advanced Quantum Studies and Department of Physics, Beijing Normal University, Beijing, China}

\author{Yue-Xin Huang}
\address{School of Sciences, Great Bay University, Dongguan, Guangdong 523000, China}
\address{Great Bay Institute for Advanced Study, Dongguan 523000, China}

\author{Xiaoxin Yang}
\address{Institute of Applied Physics and Materials Engineering, University of Macau, Taipa, Macau, China}

\author{Yi Liu}
\email{yiliu@bnu.edu.cn}
\address{Center for Advanced Quantum Studies and Department of Physics, Beijing Normal University, Beijing, China}

\author{Cong Xiao}
\email{xiaoziche@gmail.com}
\address{Institute of Applied Physics and Materials Engineering, University of Macau, Taipa, Macau, China}

\author{Zhe Yuan}
\email{yuanz@fudan.edu.cn}
\address{Institute for Nanoelectronics and Quantum Computing, Fudan University, Shanghai, China}

\begin{abstract}
Nonlinear spin polarization occurring in the second order of driving electric current is the dominant source of nonequilibrium magnetization in centrosymmetric or weakly noncentrosymmetric nonmagnetic materials, and induces nonlinear spin-orbit torque in magnets. Up to now, only the intrinsic mechanism based on anomalous spin polarizability dipole, which is the spin counterpart of Berry curvature dipole, has been studied, while disorder induced mechanisms are still missing. Here, we derive these contributions, which include not only the anomalous distribution function due to skew scattering and coordinate shift, but also interband coherence effects given by disorder induced spin shift and electric field induced anomalous scattering amplitude. We demonstrate these terms and show their importance in a minimal model. A scaling law for nonlinear current-induced spin polarization is constructed, which may help analyze experimental data in the future.
\end{abstract}
\maketitle

\renewcommand{\arraystretch}{2}

Current induced spin polarization (CISP) is a central effect in spintronics towards spin-charge conversion and electrical control of spin \cite{Review2009,Manchon2019}. In linear response to the driving electric current, the effect was originally proposed in nonmagnetic materials \cite{Pikus1978,Aronov1989,Edelstein}, can only appear in noncentrosymmetric crystals \cite{Pikus1978,Culcer2007}, and has been observed by magneto-optical means \cite{Kato2004,Stern2006}. The physics of this effect falls into the standard Boltzmann response framework \cite{Manchon2008}, and is parallel to the Drude conductivity of charge current response. Recently, spin response to the square of driving current was proposed \cite{Xiao2023NLSOT}, which can be the leading effect in nonmagnetic crystals where the inversion symmetry is maintained or not severely broken. It stems from an \textit{anomalous spin} carried by spin-orbit coupled Bloch electrons under electric field, which is determined by the momentum space dipole of anomalous spin polarizability (ASP), a geometric quantity intrinsic to the band structure. This is a Berry phase effect and is exactly the spin counterpart of the widely studied nonlinear Hall effect induced by Berry curvature dipole \cite{Fu2015,Ma2019}. 

As the nonlinear Hall effect receives significant disorder induced contributions other than the Berry curvature dipole \cite{Du2019,Kang2019,Lu2021}, one naturally asks about the role of disorder in nonlinear CISP. Moreover, it is anticipated that the interplay of the ASP-dipole intrinsic and disorder induced extrinsic contributions can be manifested via tuning system parameters such as the temperature and gate, thus some scaling law \cite{Tian2009,Hou2015,Du2019,Kang2019,Lu2021,Xiao2019scaling,Duan2022,He2022graphene,ma2022growth,Liao2023,Li2023scaling,Hueso2024} is highly desired for understanding experimental observations of nonlinear CISP. Despite the above importance, the extrinsic nonlinear CISP has not been studied. 

In this work, we develop the semiclassical theory of extrinsic contributions to the nonlinear CISP, and derive systematic formulas for different terms. The focus is on the time reversal ($\mathcal{T}$) even effect, which is allowed in both nonmagnetic and magnetic systems \cite{Zelezny2017,Xiao2023NLSOT}. We find that skew scattering and coordinate shift, which play basic roles in anomalous Hall effect \cite{nagaosa2010}, also matter for nonlinear spin response. Besides, the electric field $E$ alters the scattering amplitude, which induces interband coherence during scattering. This effect is dubbed, following the terminology of anomalous velocity that arises from field-induced interband coherence in drift motion \cite{Niu1995,Xiao2010}, as the anomalous scattering amplitude. These three mechanisms take action in the off-equilibrium electronic distribution function. In addition, the scattering potential dresses the Bloch state, leading to an interband coherence correction to the spin carried by a particular electron, i.e., a spin shift induced by scattering. We illustrate the extrinsic nonlinear CISP arising from these mechanisms in a minimal model, and find that they are in the same order of magnitude as the ASP dipole contribution \cite{Xiao2023NLSOT}. We also construct the scaling law for the phenomenon of nonlinear CISP.

\begin{figure*}[ptb]
\centering
\includegraphics[width=2\columnwidth]{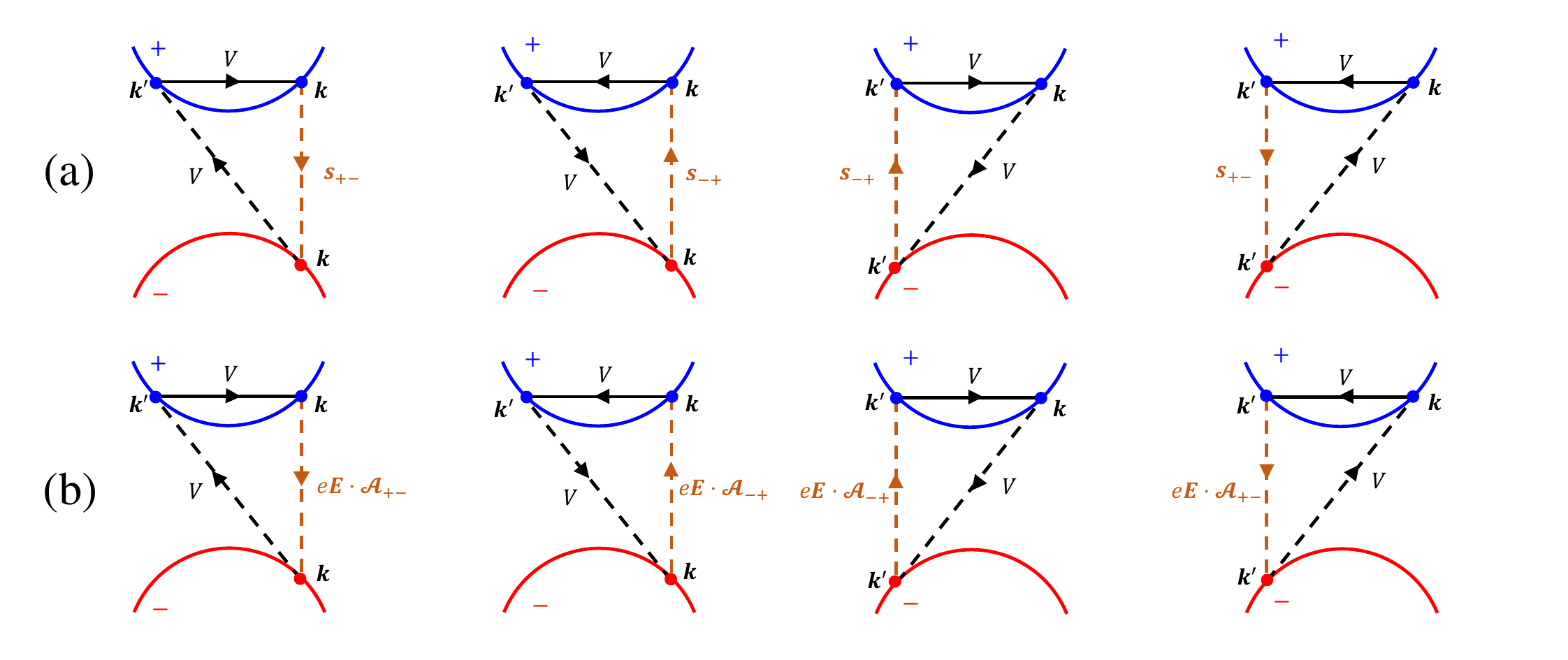}\caption{Schematics of intraband and interband transition processes involved in (a) disorder induced spin shift and (b) $E$-field induced anomalous scattering amplitude.}
\label{fig:interband}%
\end{figure*}

\emph{\color{blue}Disorder induced spin shift.}--The semiclassical response theory is a convenient tool for approaching $E$-field driven nonequilibrium phenomena. It has been successfully applied to account for extrinsic contributions in anomalous and spin Hall effects \cite{Sinitsyn2007,Sinitsyn_semiclassical_2007,xiao2018effective,xiao2018SHE}, linear spin-orbit torque \cite{Xiao2017}, as well as nonlinear Hall effects \cite{Du2019,Konig2019,Fu2020,Xiao2019NLHE,huang2023scaling}. Within the semiclassical formalism, the spin density is given by the summation of the spin polarization $\bm{s}_l$ carried by each electron weighted by the distribution function $f_l$:
\begin{equation}
\bm{S}=\frac{1}{\mathcal{V}}\sum_l f_l \bm{s}_l \label{spindensity}
\end{equation}
Here, $\mathcal{V}$ represents the volume of the system, and $l=(\eta,\bm{k})$ corresponds to the band index $\eta$ and the wave vector $\bm{k}$, respectively.

From the study of anomalous Hall effect, we know that both electric field and scattering potential can polarize the Bloch state hencealter the expectation value of an observable on that state \cite{Sinitsyn_semiclassical_2007,xiao2018effective}. Here, in the presence of scattering and $E$ field, the spin polarization carried by a particular electron is given by
\begin{equation}
\bm{s}_l=\bm{s}^0_l+\bm{s}^{\text{a}}_l+\bm{s}^{\text{ss}}_l, \label{totalspin}
\end{equation}
where $\bm{s}^0_l$ denotes the expectation value of spin operator on the unperturbed Bloch state $|{\eta\bm{k}}\rangle$, $\bm{s}^{\text{a}}_l$ and $\bm{s}^{\text{ss}}_l$ arise from field and scattering perturbed electronic state, respectively. Their expressions can be acquired by the recently developed semiclassical approach for evaluating observables other than electric current \cite{Xiao2019SHE,Xiao2017,Xiao2023NLSOT}. In particular,
\begin{equation}
    (\bm{s}^{\text{a}}_{l})_\alpha= -\frac{e}{\hbar}(\Upsilon_{l})_{\alpha \beta}E_\beta
\end{equation}
shares the same origin as the $E$-field induced anomalous velocity \cite{Xiao2010} hence is dubbed as the anomalous spin \cite{Xiao2023NLSOT}. Here and hereafter, the summation over repeated Cartesian indices $\alpha, \beta ..$ is implied. The rank-2 tensor 
\begin{equation}
\Upsilon_{\eta \bm k}=-2\hbar^2\operatorname{Im}\sum_{\eta^{\prime}\neq \eta}\frac{\bm s_{\eta \eta^{\prime}}(\bm k)\bm v_{\eta^{\prime}\eta}(\bm k)}{(\varepsilon_{\eta\bm{k}}-\varepsilon_{\eta'\bm{k}})^2} \label{geometric}%
\end{equation}
is the anomalous spin polarizability (ASP). In (\ref{geometric}),
$\varepsilon_{\eta\bm{k}}$ is the band energy, and the numerator involves the interband matrix elements of
spin and velocity operators. On the other hand, 
\begin{align}
\bm{s}^{\mathrm{ss}}_l=&-2\pi\sum_{\eta'\bm{k}'}W_{\bm k,\bm k'}^{\mathrm{o}}\delta(\varepsilon_{\eta\bm{k}}-\varepsilon_{\eta'\bm{k}'}) \nonumber\\
\times&\Im\Bigg[\sum_{\eta''\neq\eta'}\frac{\langle u_{\eta \bm{k}}|u_{\eta' \bm{k}'}\rangle \bm{s}_{\eta'\eta''}(\bm{k}') \langle u_{\eta'' \bm{k}'}| u_{\eta \bm{k}}\rangle}{\varepsilon_{\eta'\bm{k}'}-\varepsilon_{\eta'' \bm{k}'}} \nonumber\\
-&\sum_{\eta''\neq\eta} \frac{\langle u_{\eta'' \bm{k}}|u_{\eta' \bm{k}'}\rangle \langle u_{\eta' \bm{k}'}|u_{\eta \bm{k}}\rangle \bm{s}_{\eta\eta''}(\bm{k})}{\varepsilon_{\eta \bm{k}}-\varepsilon_{\eta'' \bm{k}}}\Bigg]
\label{spin shift}
\end{align}
characterizes an effective spin shift due to scattering-induced interband coherence processes~\cite{Xiao2019SHE} (see schematics in Fig. \ref{fig:interband}(a)).
Here, $W_{\bm k,\bm k'}^{\mathrm{o}}=W_{\bm k',\bm k}^{\mathrm{o}}$ is the plane-wave part of the Born scattering amplitude, and we assume scalar disorder for concreteness. In the case of static impurity, one has $W_{\bm k,\bm k'}^{\mathrm{o}}=\langle|V^{\mathrm{o}}_{\bm k,\bm k'}|^2\rangle_c$, where $\langle...\rangle_c$ indicates average over random impurity configuration, and $V^{\mathrm{o}}_{\bm k,\bm k'}$ is the plane-wave part of the scattering matrix element.

It is interesting to note that the appearance of disorder induced spin shift is also connected to the band geometric quantity ASP. This connection can be made explicit by considering scattering in the long-range limit. In this limit, in Eq. (\ref{spin shift}), $\bm k$ is very close to $\bm{k}'$ hence $\eta'$ is forced to be equal to $\eta$. Then, expanding the integrand of (\ref{spin shift}) up to the first order of $\bm{k}'-\bm k$, we get 
\begin{equation}
(\bm{s}^{\mathrm{ss}}_l)_\alpha=(\Upsilon_{l})_{\alpha \beta}\sum_{\bm{k}'}\Tilde{\omega}_{\bm{kk}'}^{(2)}(k_\beta-k'_\beta) \simeq (\Upsilon_{l})_{\alpha \beta}k_\beta/\tau,
\label{spin shift-1}
\end{equation}
where $\Tilde{\omega}_{\bm{kk}'}^{(2)}=\frac{2\pi}{\hbar}\langle|V^{\mathrm{o}}_{\bm k,\bm k'}|^2\rangle_c\delta(\varepsilon_{\eta\bm{k}}-\varepsilon_{\eta\bm{k}'})$ is the scattering rate for long-ranged disorder in the lowest Born order. The integration in (\ref{spin shift-1}) indicates the momentum relaxation $k_\beta/\tau$ within the momentum relaxation time ($\tau$) approximation.

The link between $\bm{s}^{\mathrm{ss}}_l$ and ASP makes it convenient to compare the relative importance of field induced anomalous spin and disorder induced spin shift. For electrons around the Fermi surface, the relative ratio of the two takes the form of 
\begin{equation}
    \frac{(\bm{s}^{\text{a}}_{l})_\alpha}{(\bm{s}^{\mathrm{ss}}_l)_\alpha} \sim \frac{-eE\tau}{\hbar k_{F}}.
\end{equation}
Here $-eE\tau$ measures the shift of Fermi surface in momentum space, which is usually much less than the Fermi momentum $\hbar k_{F}$ \cite{ashcroft1976solid}. 

Although $\bm{s}^{\mathrm{ss}}$ is much larger than $\bm{s}^{\text{a}}$ on the Fermi surface, their contributions to macroscopic nonlinear spin response are anticipated to be generally in the same order of magnitude. To see this, we inspect the semiclassical Boltzmann equation that describes the distribution function $f_l$ of electrons:
\begin{equation}
\frac{e}{\hbar}\bm{E}\cdot\partial_{\bm{k}}f_l=-\sum_{l'}\left(\omega_{l'l}f_l-\omega_{ll'}f_{l'}\right).\label{SBM}
\end{equation}
The right-hand side is the collision integral, where $\omega_{l'l}$ is the scattering rate from state $l$ to $l'$.
$f_l$ can be solved in ascending powers of $E$ field:
\begin{equation}
    f_l = f_{0,l}+f_{1,l}+ f_{2,l},
\end{equation}
where $f_{n,l}$ is the distribution function in the $E^n$ order. Under the relaxation time approximation, one has $f_{2,l}/f_{1,l} \sim -eE\tau/\hbar k_{F}$. Then, according to Eq. (\ref{spindensity}), the ASP and spin-shift contributions to spin response in the $E^2$ order are given by $\sum_l f_{1,l}\bm{s}^{\text{a}}_{l}$ and $\sum_l f_{2,l}\bm{s}^{\text{ss}}_{l}$, respectively, and are of the same order of magnitude. 

\begin{figure}[ptb]
\centering
\includegraphics[width=1\columnwidth]{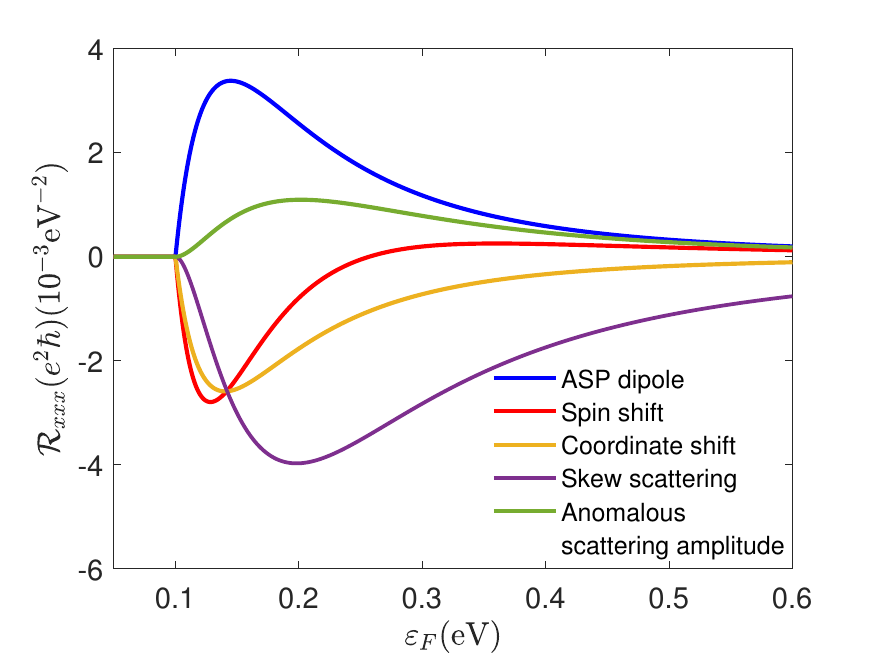}\caption{The second order nonlinear CISP of model (\ref{rashba}), plotted according to the expressions in Table. \ref{table: expressions}. The total skew scattering comes from the sum of Gaussian and non-Gaussian terms. Parameters are chosen as $w=0.1$ eV$\cdot$\AA, $v=1$ eV$\cdot$\AA, $\Delta=0.1$ eV, $n_iV_0^2=10^2$ (eV$\cdot$\AA)$^2$ and $n_iV_1^3=10^4$ eV$^3\cdot$\AA$^4$. }
\label{fig:e_f}%
\end{figure}

The above qualitative analysis shows that the extrinsic contribution to nonlinear spin from the spin shift mechanism is comparable to the intrinsic ASP term in the case of smooth disorder potential. In the later quantitative calculation on a minimal model, we show the same conclusion  for short-ranged disorder (see the blue and red curves in Fig. \ref{fig:e_f}). One can thus expect that for nonlinear spin, the extrinsic and intrinsic contributions are in general both important.

\emph{\color{blue}Field induced anomalous scattering amplitude.}--From the study of nonlinear Hall effect \cite{Lu2021}, it is known that the skew scattering and the field effect during scattering, which are beyond the above simple relaxation time approximation, can also contribute to nonlinear response by altering the distribution function. The skew scattering stems from higher-order Born expansions of the scattering rate, including the non-Gaussian conventional skew scattering and the Gaussian intrinsic skew scattering \cite{Sinitsyn_semiclassical_2007}. The field effect during scattering consists of not only the electric field working upon the coordinate shift process, which has been known for a long time as a part of the side jump mechanism for linear anomalous Hall effect \cite{Sinitsyn_semiclassical_2007,nagaosa2010}, but also the field corrected scattering amplitude, which is unique to nonlinear response and starts to contribute from the $E^2$ order \cite{Xiao2019NLHE}. 

As the skew scattering and coordinate shift are well known, the pertaining formulation is relegated to Supplemental Material \cite{supp}. As for the relatively new nonlinear contribution from field corrected scattering amplitude, it has thus far often been regarded as another kind of side jump contribution, although its physical origin and picture are not related to any side jump of electron. Here, considering its origin in the field polarized Bloch state, which also underlies the anomalous velocity and anomalous spin of Bloch electrons \cite{Xiao2023NLSOT}, we dub this contribution as the anomalous scattering amplitude. In the case of scalar disorder, the corresponding change of scattering rate in the lowest Born order is given by 
$
\omega_{l'l}^{(2),\mathrm{asa}}=\omega_{ll'}^{(2),\mathrm{asa}}=\frac{2\pi}{\hbar}W_{l'l}^{\mathrm{asa}}\delta(\varepsilon_l-\varepsilon_{l'})
$, where the anomalous scattering amplitude reads
\begin{widetext}
\begin{equation}
W_{l'l}^{\mathrm{asa}}=W_{\bm k,\bm k'}^{\mathrm{o}}(-e\bm{E})\cdot
2\Re\Bigg[\sum_{\eta''\neq\eta'} \frac{\langle u_{\eta \bm{k}}|u_{\eta' \bm{k}'}\rangle \bm{\mathcal{A}}_{\eta'\eta''}(\bm{k}') \langle u_{\eta'' \bm{k}'}|u_{\eta \bm{k}}\rangle}{\varepsilon_{\eta' \bm{k}'}-\varepsilon_{\eta'' \bm{k}'}}+\sum_{\eta''\neq\eta}\frac{\langle u_{\eta \bm{k}}|u_{\eta' \bm{k}'}\rangle\langle u_{\eta' \bm{k}'}| u_{\eta'' \bm{k}}\rangle \bm{\mathcal{A}}_{\eta''\eta}(\bm{k})}{\varepsilon_{\eta\bm{k}}-\varepsilon_{\eta'' \bm{k}}} \Bigg],
\label{asa}
\end{equation}
\end{widetext}
with $\bm{\mathcal{A}}_{\eta'\eta}(\bm{k})=\langle u_{\eta'\bm{k}}|i\partial_{\bm{k}}|u_{\eta \bm{k}}\rangle$ being the interband Berry connection. The schematics of intraband and interband transition processes involved in $W_{l'l}^{\mathrm{asa}}$ are shown in Fig. \ref{fig:interband}(b).
Comparing the $E$-field induced anomalous scattering amplitude (\ref{asa}) [Fig. \ref{fig:interband}(b)] with the disorder induced spin shift (\ref{spin shift}) [Fig. \ref{fig:interband}(a)], one observes interesting structural similarity.

One may immediately ask why this anomalous scattering amplitude has not been found in any linear response of current and spin. In fact, substituting $\omega_{l'l}^{(2),\mathrm{asa}}$ into the collision integral of Boltzmann equation (\ref{SBM}), one finds that the pertaining term vanishes at the first order of $E$ field. This means that the contribution from anomalous scattering amplitude is a purely nonlinear response phenomenon. In particular, in the $E^2$ order, the anomalous scattering amplitude gives rise to an effective driving term in the Boltzmann equation, which reads
\begin{equation}
    \sum_{l'}\omega_{l'l}^{(2),\mathrm{asa}}(f_{1,l}-f_{1,l'})=-\sum_{l'}\omega_{l'l}^{(2)}(f_{2,l}^{\mathrm{asa}}-f_{2,l'}^{\mathrm{asa}}).
\end{equation}
The solution of this equation, $f_{2,l}^{\mathrm{asa}}$, yields an additional distribution function in the $E^2$ order.


\begin{table*}[pbt]
\begin{centering}
\caption{Mechanisms and expressions for extrinsic contributions to the nonlinear CISP response tensor of model (\ref{rashba}). Short-range random scalar impurity potential $V(\boldsymbol{r})=\sum_iV_i\delta(\bm r - \bm R_i)$ is considered, with $\langle V_i\rangle_c=0$, $\langle {V}_{i}^2\rangle_c=V_0^2$ and $\langle {V}_{i}^3\rangle_c=V_1^3$, and $n_i$ is the density of impurities.}
\label{table: expressions}
\begin{tabularx}{\linewidth}{@{\hspace{4em}} l@{\hspace{10em}} X}
\hline\hline
\multicolumn{1}{@{\hspace{7em}} l}{Mechanism} & \multicolumn{1}{@{\hspace{2em}} l}{Nonlinear response coefficient}\\
\hline
Anomalous spin  polarizability dipole & $\mathcal{R}_{xxx}^{\text{asp}}=-\frac{3 e^2 \hbar v w \Delta \left(\Delta ^2-\varepsilon _F^2\right)}{2 \pi n_iV_0^2 \varepsilon _F^3 \left(3 \Delta ^2+\varepsilon _F^2\right)}$  \\
Spin shift    &  $\mathcal{R}_{xxx}^{\text{ss}}=\frac{e^2\hbar  v w   \Delta  \left(\Delta^2 -\varepsilon _F^2\right) \left(33 \Delta ^2-5 \varepsilon _F^2\right)}{4 \pi  n_iV_0^2 \varepsilon _F^3 \left(3 \Delta ^2+\varepsilon _F^2\right)^2}$   \\
Coordinate shift  &  $\mathcal{R}_{xxx}^{\text{cs}}=\frac{  e^2\hbar v w \Delta \left(\Delta^2 -\varepsilon _F^2\right)  \left(17 \Delta ^2+3 \varepsilon _F^2\right)}{4 \pi  n_iV_0^2 \varepsilon _F^3 \left(3 \Delta ^2+\varepsilon _F^2\right)^2}$ \\
Anomalous scattering amplitude  &  $\mathcal{R}_{xxx}^{\text{asa}}=\frac{3e^2 \hbar v w  \Delta   \left(\Delta ^2-\varepsilon _F^2\right)^2}{2 \pi  n_iV_0^2 \varepsilon _F^3 \left(3 \Delta ^2+\varepsilon _F^2\right)^2}$  \\
Conventional skew scattering  &   $\mathcal{R}_{xxx}^{\text{csk}}=-\frac{e^2 \hbar v w V_1^3   \Delta\left(\Delta ^2-\varepsilon_{F}^2\right)^2 \left(9 \Delta ^2+5 \varepsilon_{F}^2\right)}{ \pi  \varepsilon_{F}^2 n_i^2V_0^6 \left(3 \Delta ^2+\varepsilon_{F}^2\right)^3}$ \\
Gaussian skew scattering     &   $\mathcal{R}_{xxx}^{\text{Gsk}}=-\frac{e^2\hbar v w  \Delta    \left(\Delta ^2-\varepsilon_{F}^2\right)^2 \left(77 \Delta ^2+13 \varepsilon_{F}^2\right)}{4 \pi  n_iV_0^2  \varepsilon_{F}^3\left(3 \Delta ^2+\varepsilon_{F}^2\right)^3}$ \\
\hline\hline
\end{tabularx}
\par\end{centering}
\end{table*}

\emph{\color{blue}Model Calculation.}--Gathering the aforementioned ingredients, one can get several extrinsic contributions to the $\mathcal T$-even nonlinear CISP by Eq. (\ref{spindensity}), with the detailed formulation presented in \cite{supp}. 
To illustrate these contributions, we apply the theory to a four-band $\boldsymbol{k}\cdot\boldsymbol{p}$ model with inversion symmetry
\begin{align}
    H(\boldsymbol{k})= w k_x \sigma_z+v(k_x s_y-k_y s_x) \sigma_z
    +\Delta s_z,
    \label{rashba}
\end{align}
where $s_{i}$'s and $\sigma_i$'s are the Pauli matrices representing the spin and orbital degrees of freedom, respectively; $\boldsymbol{k}= (k_x,k_y)$ is the wave vector; $w$, $v$ and $\Delta$ are the model parameters. $w$ tilts the Dirac cone along the $x$ direction. This model consists of two copies of tilted Weyl model connected by the inversion operation. It is defined around one valley in the Brillouin zone of nonmagnetic systems, whereas its time-reversed counterpart can be written down for the other valley. As we are considering a $\mathcal T$-even response, the existence of $\mathcal T$-connected two valleys simply doubles the result. Furthermore, regarding symmetry, the presence of inversion forbids the linear CISP, and the $\mathcal{M}_x\mathcal T$ symmetry of $H(\boldsymbol{k})$ ensures that for the second order nonlinear response defined by 
\begin{equation}
    S_\alpha=\mathcal{R}_{\alpha\beta\gamma}
    E_{\beta}E_{\gamma},
\end{equation}
only $\mathcal{R}_{xxx}$, $\mathcal{R}_{xyy}$, $\mathcal{R}_{y(xy)}$, and $\mathcal{R}_{z(xy)}$ are allowed. Here $\mathcal{R}_{z(xy)}\equiv (\mathcal{R}_{zxy}+\mathcal{R}_{zyx})/2$. For illustrative purpose, we calculate $\mathcal{R}_{xxx}$ in the following.

We consider short-range random impurity potential $V(\boldsymbol{r})=\sum_iV_i\delta(\bm r - \bm R_i)$, with $\langle V_i\rangle_c=0$, $\langle {V}_{i}^2\rangle_c=V_0^2$ and $\langle {V}_{i}^3\rangle_c=V_1^3$, and solve the Boltzmann equation with skew scattering, coordinate shift and anomalous scattering amplitude, following the method and approximations adopted in 
~\cite{Du2019,Konig2019,Fu2020,Xiao2019NLHE}. In particular, to obtain analytic result, we assume $w\ll v$. The expressions of $\mathcal{R}_{xxx}$ induced by different semiclassical mechanisms are shown in Table \ref{table: expressions}. The calculation details are provided in \cite{supp}. One sees that the common factor $(\Delta ^2-\varepsilon_{F}^2)$ in the numerator of all contributions ensures the vanishing of each term at the band edge. The tilt term $w$ is crucial, because it breaks out-of-plane rotational axis that would otherwise forbid the in-plane response. In Fig.~\ref{fig:e_f}, we plot different contributions as a function of the Fermi energy, and find them to be comparable in magnitude.

\emph{\color{blue}Scaling law of the nonlinear CISP.}--Because of the co-existence of nonlinear CISP from multiple origins, it is helpful to have some guidelines for understanding experimental data. In this regard, the scaling law between the detected spin signal and longitudinal resistivity (conductivity) may render useful information, which has been shown in anomalous and spin Hall effects \cite{Hou2015}, spin-orbit torque \cite{Manchon2019}, as well as nonlinear Hall effect \cite{Kang2019,Du2019}. Given the parallel formulation of nonlinear CISP and nonlinear Hall effect, they should possess the same scaling law (one can readily check this). For instance, in the presence of two types of static disorder, in which one is impurity ($i=0$) and the other is phonon ($i=1$) \cite{Hou2015,Du2019}, one has
\begin{equation}
\mathcal{R} \rho=C+A_{0}\frac{\rho_{0}}{\rho^{2}}%
+\sum_{i=0,1}C_{i}\frac{\rho_{i}}{\rho}+\sum_{i,j=0,1}C_{ij}\frac
{\rho_{i}\rho_{j}}{\rho^{2}},
\end{equation}
where $\rho=\rho_0+\rho_1$ is the longitudinal resistivity, and $\rho_0$ is the residual resistivity. The scaling parameter $C$ stands for the ASP dipole contribution, $A_0$ and $C_{ij}$ are from conventional and Gaussian skew scattering, respectively, and $C_i=C_{i}^{\mathrm{ss}}+C_{i}^{\mathrm{cs}}+C_{i}^{\mathrm{asa}}$ from the spin shift, coordinate shift as well as the anomalous scattering amplitude. The scaling law can also be expressed in terms of longitudinal conductivity ($\sigma \simeq 1/\rho, \sigma_0 \simeq 1/\rho_0$)
\begin{equation}
\mathcal{R}/\sigma-A_{0}\sigma^{2}/\sigma_{0}=B+B^{\prime}%
\sigma/\sigma_{0}+B^{\prime\prime}\left(  \sigma/\sigma_{0}\right)
^{2}, \label{scaling}%
\end{equation}
where $B=C+C_{1}+C_{11}$, $B^{\prime}=C_{0}-C_{1}+C_{01}+C_{10}-2C_{11}$ and
$B^{\prime\prime}=C_{00}+C_{11}-C_{01}-C_{10}$. 
Note that the static approximation of electron-phonon scattering is practically valid as long as $\rho$ has a nearly linear temperature dependence \cite{Xiao2019scaling}, and this behavior usually extends to quite low temperatures in moderately disordered samples fabricated in most spintronics experiments \cite{Manchon2019}.

At low temperatures where $\sigma \simeq \sigma_0$, scaling (\ref{scaling}) becomes $\mathcal{R}/\sigma=A_{0}\sigma_{0}+C+C_{0}+C_{00}$, by which $A_0$ can be determined in experiments. Then, at finite temperatures, $\mathcal{R}/\sigma-A_{0}\sigma^{2}/\sigma_{0}$ can be fitted as a parabolic function of $\sigma/\sigma_{0}$. Noticeably, the linear and quadratic terms in this fitting can only arise from extrinsic mechanisms.

\emph{\color{blue}Discussion.}--We have shown the importance of extrinsic contributions to nonlinear CISP by semi-quantitative analysis and quantitative model calculations. We highlight the nonlinear responses from disorder induced spin shift and field induced anomalous scattering amplitude, which have received little attention in previous studies of spintronics and nonlinear electronics. The proposed scaling law is expected to serve as a first step to understand the interplay of intrinsic and extrinsic contributions in experimental data.

Besides nonlinear CISP, the $\mathcal{T}$-even nonlinear current induced orbital magnetization has also received recent interest \cite{Lee2024}. Despite the complexity of accurately formulating the nonequilibrium orbital magnetization in metals introduced by the nonlocality of orbital magnetic dipole operator \cite{Xiao2021OM}, the scaling law as a qualitative result should still be the same as that for nonlinear CISP.

In this work we focused on $\mathcal{T}$-even nonlinear CISP, thus the scaling law obtained is the same as that for $\mathcal{T}$-even nonlinear charge current response \cite{Du2019}. In magnetic systems, $\mathcal{T}$-odd nonlinear CISP can also occur \cite{Wang2022,Xiao2022NLSOT}, and the pertinent scaling law is more involved, which is the same as that for $\mathcal{T}$-odd nonlinear charge transport \cite{huang2023scaling}.

\emph{Acknowledgements.} This work was supported by the National Natural Science Foundation of China (Grant No. 12174028 and No. 12374101), and UM Start-up Grant (SRG2023-00033-IAPME).

\bibliography{ref}

\end{document}